\documentclass[letter]{aa}  

\usepackage{graphicx}
\usepackage{txfonts}

\usepackage[breaklinks,colorlinks,citecolor=blue,linkcolor=blue,urlcolor=blue]{hyperref}
\begin{document}

   \title{The VMC survey}
   
   \subtitle{XXX. Stellar proper motions in the central parts of the Small Magellanic Cloud\thanks{Based on observations made with VISTA at the Paranal Observatory under program ID 179.B-2003}}

   \author{F. Niederhofer
          \inst{1}
          \and
          M.-R. L. Cioni
          \inst{1}
          \and
          S. Rubele
          \inst{2,3}
          \and
          T. Schmidt
          \inst{1}
          \and
          K. Bekki
          \inst{4}
          \and
          R. de Grijs
          \inst{5,6}
          \and
          J. Emerson
          \inst{7}
          \and
          V. D. Ivanov
          \inst{8,9}
          \and
          M. Marconi
          \inst{10}
          \and
          J. M. Oliveira
          \inst{11}
          \and
          M. G. Petr-Gotzens
          \inst{9}
          \and
          V. Ripepi
          \inst{10}
          \and
          J.~Th.~van~Loon
          \inst{11}
          \and
          S. Zaggia
          \inst{2}
          }

   \institute{Leibniz-Institut f\"ur Astrophysik Potsdam, An der Sternwarte 16, D-14482 Potsdam, Germany
              \and
              Osservatorio Astronomico di Padova $-$ INAF, Vicolo dell'Osservatorio 5, I-35122 Padova, Italy
              \and
              Dipartimento di Fisica e Astronomia, Universit{\'a} di Padova, Vicolo dell’Osservatorio 2, I-35122 Padova, Italy
              \and
              ICRAR, M468, University of Western Australia, 35 Stirling Hwy, 6009 Crawley, Western Australia, Australia
              \and
              Department of Physics and Astronomy, Macquarie University, Balaclava Road, Sydney, NSW 2109, Australia
              \and
              International Space Science Institute--Beijing, 1 Nanertiao, Zhongguancun, Hai Dian District, Beijing 100190, China
              \and
              Astronomy Unit, School of Physics and Astronomy, Queen Mary University of London, Mile End Road, London E1 4NS, UK
              \and
              European Southern Observatory, Ave. Alonso de Cordova 3107, Vitacura, Santiago, Chile
              \and
              European Southern Observatory, Karl-Schwarzschild-Str. 2, D-85748 Garching bei M\"{u}nchen, Germany
              \and
              INAF-Osservatorio Astronomico di Capodimonte, via Moiariello 16, 80131, Naples, Italy
               \and
             Lennard-Jones Laboratories, 
             Keele University, ST5 5BG, UK
             }

   \date{Received; accepted}

 
  \abstract
  {
  We present the first spatially resolved map of stellar proper motions within the central ($\sim$3.1 $\times$ 2.4~kpc) regions of the Small Magellanic Cloud (SMC). The data used for this study encompasses four tiles from the ongoing near-infrared VISTA survey of the Magellanic Clouds system and covers a total contiguous area on the sky of $\sim$6.81~deg$^2$. Proper motions have been calculated independently in two dimensions from the spatial offsets in the $K_s$ filter over time baselines between 22 and 27 months. The reflex motions of $\sim$33~000 background galaxies are used to calibrate the stellar motions to an absolute scale. The resulting catalog is composed of more than 690 000 stars which have been selected based on their position in the $(J-K_s, K_s)$ color-magnitude diagram. For the median absolute proper motion of the SMC, we find ($\mu_{\alpha}\mathrm{cos}(\delta)$, $\mu_{\delta}$) = (1.087~$\pm$~0.192 (sys.)~$\pm$~0.003 (stat.), $-$1.187~$\pm$~0.008 (sys.)~$\pm$~0.003 (stat.))~mas~yr$^{-1}$, consistent with previous studies. Mapping the proper motions as a function of position within the SMC reveals a non uniform velocity pattern indicative of a tidal feature behind the main body of the SMC and a flow of stars in the South-East moving predominantly along the line-of-sight.
}


   \maketitle
%

\section{Introduction \label{sec:intro}}

The Large and the Small Magellanic Cloud (LMC and SMC) are the most prominent dwarf galaxy satellites of the Milky Way and are in the early phases of a minor merger event. Tidal interactions between the two Clouds led to the formation of the gaseous Magellanic Stream \citep[see e.g.,][]{Nidever08, D'Onghia16} and the Magellanic Bridge \citep[e.g.][]{Irwin85} which is composed of stars and gas, and connects the two dwarf galaxies. Thanks to their close vicinity, the Clouds provide a unique opportunity for studying in detail the stellar kinematics within an interacting pair of galaxies. However, for a long time, the one dimensional line-of-sight velocities of stars have been the only source of dynamical information. With accurate proper motion measurements, which are now available using modern observing facilities, both ground and space based, it is now possible to assess the full three-dimensional velocity field of stellar populations, both as a function of age and position within the Magellanic Clouds. Measurements of stellar proper motions using the \textit{Hubble Space Telescope} (HST) have changed the traditional picture that the Magellanic Clouds have orbited the Milky Way several times. \citet{Kallivayalil06b,Kallivayalil06a,Kallivayalil13} showed that both satellites move faster around our Galaxy than originally thought. This implies that the Clouds are rather on a first passage or a long period orbit, depending on the mass of the Milky Way and the LMC \citep[e.g.,][]{Besla07,Patel17}. The internal proper motion field of the LMC, which was measured by \citet{vanderMarel14} for the first time, reveals a clockwise rotation of the LMC disc in the plane of the sky, consistent with previous line-of-sight velocity measurements. This result was confirmed by \citet{vanderMarel16} using the proper motions from the \textit{Tycho-Gaia} Astrometric Solution (TGAS) Catalog \citep{Michalik15,Lindegren16}. The SMC, on the other hand, seems to have a more complex kinematical structure. Radial velocity measurements of intermediate-age red giant branch (RGB) stars show signs of tidal stripping in the outer regions of the SMC and a velocity gradient along the Northwest-Southeast axis \citep{Dobbie14}. The young stellar populations also show such velocity gradient, as well as a higher velocity towards the SMC Wing, a horizontal extension of the SMC towards the East \citep{Evans08}.

Here we present for the first time a large scale map of stellar proper motions within the central $\sim3.1\times2.4$~kpc of the SMC, covering a total contiguous sky area of $\sim$6.81~deg$^2$. The data for this study stem from the multi-epoch near-infrared VISTA survey of the Magellanic Clouds system \citep[VMC,][]{Cioni11}. Our group has already analyzed some individual VMC tiles to measure the proper motions of stars within the LMC \citep{Cioni14}, SMC \citep{Cioni16} and the Galactic globular cluster 47~Tuc \citep{Cioni16,Niederhofer18}.

\section{Observations and photometry \label{sec:obs}}

The data used in this study are taken with the Visible and Infrared Survey Telescope for Astronomy \citep[VISTA,][]{Sutherland15} and comprise of four VISTA tiles, namely SMC 4\_3, 4\_4, 5\_3 and 5\_4. Each tile is composed of six individual pawprint images with specific offsets, in order to observe a continuous area on the sky, owing to the gaps between the 16 VIRCAM detectors in the field of view \citep{Emerson06}. In the stacked tile, each object is observed at least twice, whereas a small fraction can be observed up to six times. In two narrow stripes at the edge of a tile, sources are only observed once. Each tile corresponds to 1.77~deg$^2$ on the sky. The tiles that cover the SMC are arranged such that stripes of single observations within a tile overlap with adjacent tiles in a North-South direction. Therefore, the area covered by the four VMC tiles is $\sim$6.81~deg$^2$. The VMC survey plan includes multiple observations of each VMC tile, with 3 epochs in the $Y$ and $J$ band and 12 epochs in the $K_s$ filter (whereas one epoch per filter is split into two shallow exposures), spread over a mean period of two years. For the calculation of the proper motions within the four SMC tiles, we use only the observations in the $K_s$ filter (central wavelength
2.15 $\mu$m) to minimise problems with differential atmospheric diffraction. We have time baselines of 23 months for SMC~4\_3, 27 months for SMC~4\_4, 22 months for SMC~5\_3 and 24 months for SMC~5\_4. Details about the tiles used here are given in Table~\ref{tab:tiles}. The astrometry of VISTA pawprints shows a systematic pattern of the order of 10$-$20 mas coming from residual World Coordinate System errors. This effect limits the precision of proper motion measurements of single objects but our overall results are in good agreement with recent studies suggesting that there is no large systematic offset in the astrometry.

For the proper motion calculations we used pawprint images at individual epochs which were reduced and calibrated with the VISTA Data Flow System (VDFS) pipeline v1.3 \citep{Irwin04, Gonzalez18} and retrieved from the VISTA Science Archive\footnote{\url{http://horus.roe.ac.uk/vsa}} \citep[VSA,][]{Cross12}. We performed point spread function (PSF) photometry on each of these images separately, as described by \citet{Rubele15}. We also created a deep multi-band catalog which will be used in the following. For this, we performed PSF photometry on deep tile images where individual pawprints from all epochs were homogenized to have a constant reference PSF and then combined to a single tile \citep[see][]{Rubele15}. Finally, we cross-matched the catalogs in all bands using a 1\arcsec ~matching radius.

\begin{table*} \small
\centering
\caption{Overview of the VMC tiles used for this study and the proper motion results. \label{tab:tiles}}
\begin{tabular}{@{}l@{ }c@{ }c@{ }c@{ }c@{ }c@{ }c@{ }c@{ }c@{ }c@{ }}
\hline\hline
\noalign{\smallskip}
Tile & \multicolumn{2}{c}{Central coordinates} & ~~Position Angle~~  & ~~Number~~ & ~~Time baseline~~ & ~~~~Number~~~~ & ~~~~~~$\mu_{\alpha}\mathrm{cos}(\delta)$~~~~~~ & ~~~~~~$\mu_{\delta}$~~~~~~ \\
&~~~RA$_{\mathrm{J2000}}$ (deg)~~~&~~~Dec$_{\mathrm{J2000}}$ (deg)~~~& (deg) &of epochs& (months) & of SMC stars & ~~~~~(mas~yr$^{-1}$)~~~~~ &~~~~~ (mas~yr$^{-1}$)~~~~~\\
\noalign{\smallskip}
\hline
\noalign{\smallskip}
SMC~4\_3 &11.3112&$-$73.1198& $-$1.1369 & 11 &23 & ~~641~397 & 0.537 $\pm$ 0.101&$-$1.172 $\pm$ 0.014\\
SMC~4\_4 &16.3303&$-$73.0876 & $+$3.6627 & 11 & 27 & ~~429~048 &1.201 $\pm$ 0.174& $-$0.680 $\pm$ 0.029\\ 
SMC~5\_3 &11.2043&$-$72.0267& $-$1.2392 & 12 & 22 & ~~220~678 & 1.761 $\pm$ 0.304&$-$1.724 $\pm$ 0.054\\
SMC~5\_4 &16.1088&$-$71.9975& $+$3.4514 & 13 & 24 & ~~295~996 & 1.625 $\pm$ 0.325&$-$1.508 $\pm$ 0.010\\

\noalign{\smallskip}
\hline
\noalign{\smallskip}
Total & & & & & & 1~587~119 & 1.087 $\pm$ 0.192& $-$1.187 $\pm$ 0.008\\
\noalign{\smallskip}
\hline

\end{tabular}

\tablefoot{
The uncertainties refer to the systematic errors, the statistical errors are of the order of 0.003$-$0.009 mas~yr$^{-1}$ per tile in both directions.
}

\end{table*}

\section{The proper motion field of the SMC \label{sec:PM}}

We calculate the proper motions of stars within the four VMC tiles covering the central parts of the SMC for each detector and pawprint separately to minimize systematic effects when combining different pawprints. The method to determine the proper motions is described in detail by \citet{Niederhofer18}. We followed these steps for each tile separately. Briefly, we first cross-correlated the individual detection catalogs from the different epochs with the deep multi-band catalog using a matching radius of 0$\farcs$5 to remove spurious detections. We then identified sources that are most likely background galaxies  based on several selection criteria involving the position in the color-color diagram and the sharpness of the source \citep[see][]{Niederhofer18}. We also selected only well measured galaxies with photometric uncertainties of $<$0.1 mag in the $K_s$ filter. Our final list of background galaxies within all four tiles contains $\sim$54~900 galaxies. Next, we transformed the $x$ and $y$ detector positions of the sources in the catalogs from all epochs to a common reference frame in order to compensate for small pointing differences. As the reference epoch, to which all other epochs were transformed, we chose the ones with the best seeing conditions (between 0$\farcs$76 and 0$\farcs$83). We performed the transformation in two stages. First, we used the sample of galaxies, which are assumed to be at rest, to perform an initial transformation and then used the stars themselves for a more refined transformation. For both steps we chose a general fit geometry, allowing for a shift, rotation and scaling in two directions. After inspection of the results of the transformations, we decided to remove two epochs for tiles SMC~4\_3 and SMC~4\_4 and one epoch for tile SMC~5\_3 
since these epochs had systematically higher residuals in the transformation, most likely due to larger seeing or airmasses during the observations. We calculated the proper motions of the stars and the reflex motions of the galaxies by fitting a linear least-squares regression model independently to the $x$ and $y$ positions of each source on the detector chip which has been detected in all epochs as a function of the Mean Julian Date. The resulting slope gives the proper motion of the source in pixels~day$^{-1}$ which was then transformed to mas~yr$^{-1}$. Finally, we corrected the motions of the stars by the mean residual reflex motion of the galaxies within each tile, to put the proper motions on an absolute scale. Our final proper motion catalogs consist of $\sim$1~770~000 stars and $\sim$33~000 galaxies of which $\sim$760~000 and $\sim$17~000, respectively, are unique sources.

\begin{figure}
\centering
\includegraphics[width=\columnwidth]{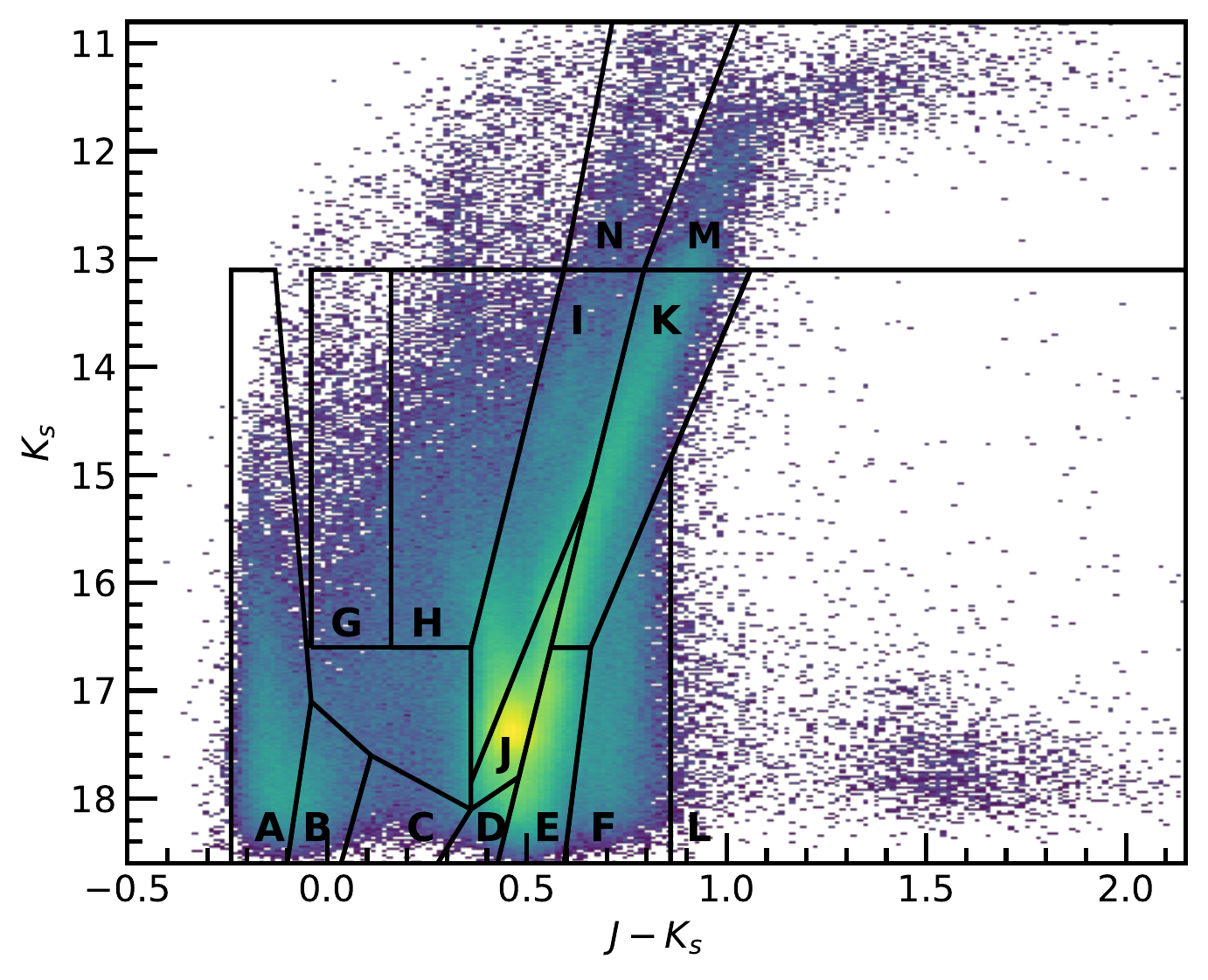}
\caption{Stellar density (Hess) diagram in the $(J-K_s, K_s)$ color-magnitude space of all sources with proper motions measurements. Overplotted as black polygons are regions of different stellar populations as defined by \citet{Cioni16} and El~Youssoufi et al. (in prep.), the latter adding regions M and N and extending regions A and L.
}
\label{fig:hess}
\end{figure}

\begin{figure*}
\centering
\includegraphics[width=18.9cm]{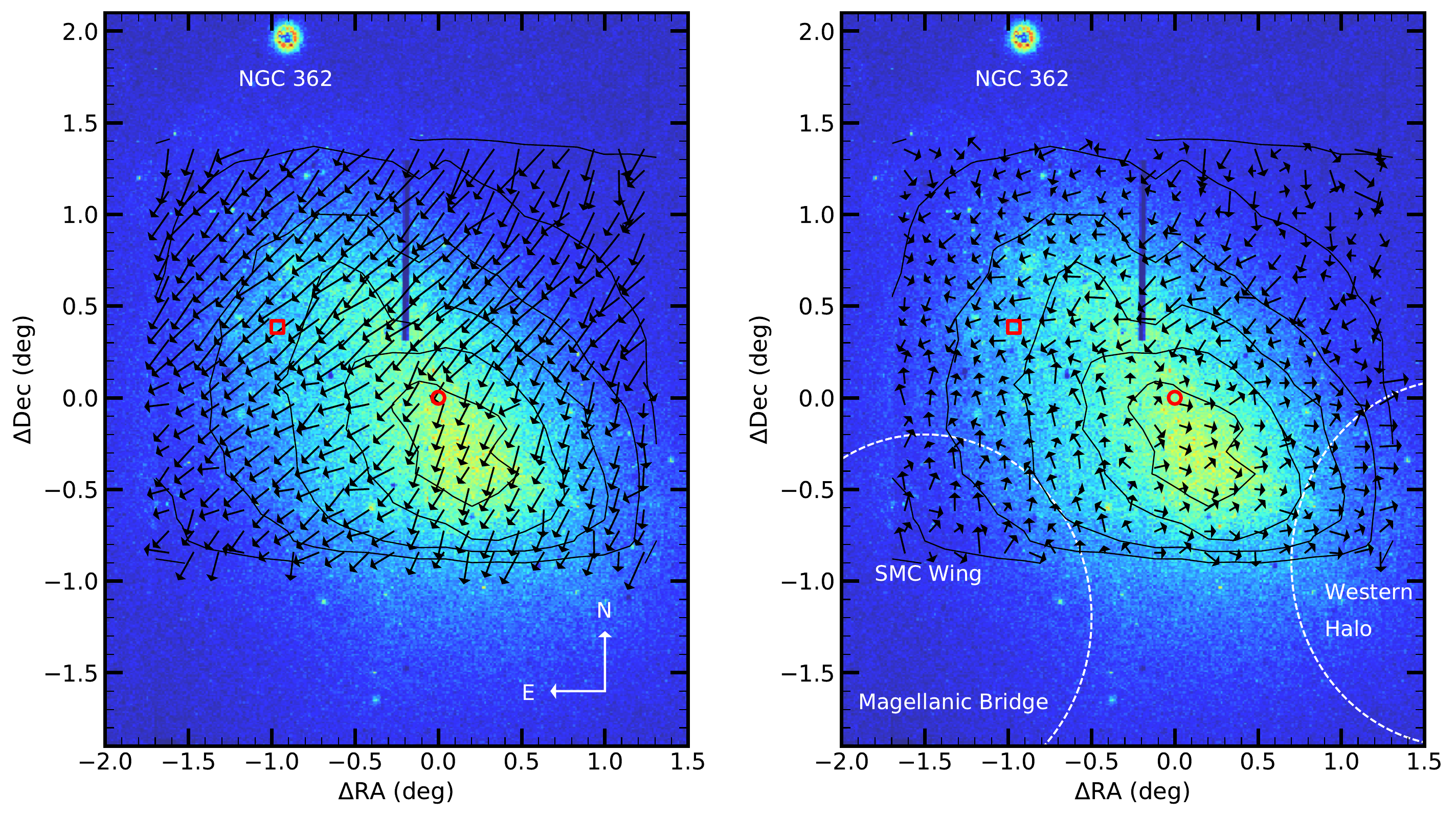}
\caption{Proper motion field (black arrows) of the central regions of the SMC from four tiles of VMC data. The background color image is a density map of objects detected by the VMC. The panel is centered at the optical center of the SMC (red circle) as given by \citet{deVaucouleurs72}, also shown as a red square is the kinematical \ion{H}{i} center \citep{Stanimirovic04}. 
North of the main body of the SMC, the foreground Galactic globular cluster NGC~362 is visible. Black solid lines show density contours of sources for which proper motions have been measured. The contours are at (500, 1500, 3500, 5500, 7500 and 9500) stars per grid cell. The vertical dark stripe at $\Delta$RA$\sim-$0.2$\degr$ is due to a narrow gap in the observations. In the left-hand panel, the arrows indicate the observed absolute proper motion, whereas in the right-hand panel, the arrows show the residual proper motions after subtraction of the systemic velocity of the SMC.}
\label{fig:pm_map}
\end{figure*}

To get a cleaner sample of stars belonging only to the SMC, we selected stars that fall within various regions in the $(J-K_s, K_s)$ color-magnitude diagram (CMD) which are dominated by SMC populations (see Figure~\ref{fig:hess}). Specifically, we selected stars within regions A and B (young main sequence stars), E (lower RGB stars), G (supergiants), I and N (red supergiants), J (red clump stars), K (upper RGB stars) and M (asymptotic giant branch stars). We further constrained our sample 
by selecting stars with photometric errors $\sigma(K_s)\leq0.05$~mag. Using these selections, our final SMC proper motions catalog contains $\sim$1~590~000 entries ($\sim$690~000 individual sources). We find a median proper motion of the SMC of ($\mu_{\alpha}\mathrm{cos}(\delta)$, $\mu_{\delta}$) = (1.087~$\pm$~0.192, $-$1.187~$\pm$~0.008)~mas~yr$^{-1}$ which is consistent within the uncertainties with previous ground-based measurements \citep{Vieira10,Costa11} but larger in the RA direction compared to space-based \citep{Kallivayalil13, vanderMarel16} measurements. The quoted uncertainties are the systematic errors which we estimated from the median deviations of the stellar proper motions from zero before correcting for the galaxies' reflex motion. 
We note that the uncertainties in the RA direction are considerably larger than the ones in the Dec direction. This can be explained by the fact that also Milky Way stars have been used for the transformation to a common reference frame. The proper motions of these stars are comparable to the ones in the SMC in the Dec direction but larger by a factor of $\sim5-10$ in the RA direction. Owing to the vast numbers of stars, the statistical uncertainty is of the order of 0.003~mas~yr$^{-1}$ in both directions.

For a spatially resolved map of proper motions within the SMC, we divided the total area covered by the four VMC tiles into a 20$\times$20 grid (bin size$\sim$18~700~pc$^2$) and calculated the median proper motion within each grid cell. Figure~\ref{fig:pm_map} shows the resulting map, where the vectors represent the direction and magnitude of the median motion within each bin. A stellar density map of sources detected by the VMC survey is also displayed in the background. The map is centered at the optical center of the SMC ($\alpha_{2000}=00^{\mathrm{h}}52^{\mathrm{m}}12\overset{\mathrm{s}}.5$, $\delta_{2000}=-72\degr49\arcmin43\arcsec$, \citealt{deVaucouleurs72}). Two different patterns of motion are evident from the overall velocity map. Stars in the outer regions of the SMC to the North and Southeast uniformly move towards the Southeast, whereas the velocity vector of stars located in the main body of the SMC, where the stellar density is highest, has a significantly ($>$3$\sigma$) smaller component in the RA direction (see also Table~\ref{tab:tiles}) and is therefore oriented more towards the South. Additionally, there seems to be a decrease in the absolute value of the velocity when going towards the Magellanic Bridge region to the South-East of the SMC.

\section{Discussion and conclusions \label{sec:discussion}}

\citet{Diaz12} presented dynamical \textit{N}-body models of the SMC to study in detail the kinematical history and the tidal features that arise from the interaction with the LMC and the Milky Way. 
Their simulations suggest that the formation of the Magellanic Bridge was accompanied by the appearance of another tidal feature which was named ``Counter Bridge'' by the authors. 
In their model, this new feature is expected to originate behind the center of the SMC, where we find proper motions with a small RA component,
and follow an arc-like structure, whereas most of the Counter Bridge is aligned with the SMC along the line-of-sight. The exact course of the Counter Bridge, however, is still debated but it likely extends towards the North-East of the SMC \citep[][]{MullerBekki07}.
The Counter-Bridge can reveal itself as an elongation of the SMC in the radial direction up to about 85~kpc. \citet{Ripepi17} argued that the three-dimensional distribution of Classical Cepheids within the SMC is elongated and consistent with the model by \citet{Diaz12}. The different proper motion signature in regions of highest stellar densities, compared to those in the outer regions of the SMC, can be interpreted in the light of the Counter Bridge. In Figure~\ref{fig:pm_map}, the resulting vectors are the median motions of individual stars within each bin. This means that the proper motion signal from stars originating from the Counter Bridge is superimposed on the one from stars associated with the main body of the SMC.
For the proper motion vectors to be almost vertical, a contribution of stars with motions in the RA direction opposite to the bulk motion of the SMC is required (see right-hand panel in Figure~\ref{fig:pm_map}). 
Our results would then imply that the stars at the base of the Counter Bridge first move towards the West before they turn towards the North-East.
Another possibility would be that the stars which move predominately towards the West are associated with the Western Halo, a tidal feature in the South-West of the SMC, suggested by \citet{Dias16}.

An alternative scenario is that the parts of the SMC with the highest stellar density have motions intrinsically different from the outer regions which might be in the process of being tidally stripped from the SMC. This model seems to be supported by the recent results from \citet{vanderMarel16}, whose sample of TGAS stars, after subtraction of the center-of-mass motion of the SMC, suggest that the stars in the densest regions systematically move towards the West (cf. their figure~1) whereas stars in the Eastern outskirts predominantly move towards the East. This might, however, be a selection effect, since in their study only five of eight stars in total are located in the Western part of the SMC. Individual precise proper motion measurements of a larger sample of stars will help discriminate between these two theories. We also note that the fields used to calculate the mean motion of the SMC using HST data are located in the transition region in our map, which might explain the lower value in the RA direction found by \citet{Kallivayalil13}.

In the proper motion map presented here, we also see a decrease in the absolute value of the stellar velocity in the Southeast of the SMC, towards the Wing and the Magellanic Bridge. In these regions, \citet{Evans08} and \citet{Dobbie14} found structures with high positive line-of-sight velocities, indicative of tidal stripping. This might suggest that the lower proper motion is only a projection effect and the three-dimensional velocity vector of the stars in these regions is more aligned with the line-of-sight.

In this study we have presented for the first time a detailed map of stellar proper motions within the central parts of the SMC, showing that the stars do not follow a uniform velocity pattern. In future works, we will extend this study to all VMC tiles covering the SMC, analyze the motions of different stellar populations, and compare the observational results with dynamical \textit{N}-body models to obtain a more complete picture of the stellar motions across the entire SMC.

\begin{acknowledgements}
We thank the Cambridge Astronomy Survey Unit  and the Wide Field Astronomy Unit in Edinburgh for providing calibrated data products under the support of the Science and Technology Facility Council. 

This project has received funding from the European Research Council under European Union's Horizon 2020 research and innovation programme (grant agreement No 682115).

This research made use of Astropy, a community-developed core Python package for Astronomy \citep{Astropy13}.

This study is based on observations obtained with VISTA at the Paranal Observatory under program ID 179.B-2003.

We thank the referee for useful comments and suggestions that helped improve the paper.

\end{acknowledgements}

%
   \bibliographystyle{aa} 
   \bibliography{references} 
%

\end{document}